\documentclass[%
 reprint,
 amsmath,amssymb,
 aps,
prb,
]{revtex4-2}

\usepackage[dvipdfmx]{graphicx}
\usepackage{dcolumn}
\usepackage{bm}

\begin{document}

\title{Phase transitions between helices, vortices, and hedgehogs driven by spatial anisotropy in chiral magnets}

\author{Kotaro Shimizu, Shun Okumura, Yasuyuki Kato, and Yukitoshi Motome}

\affiliation{Department of Applied Physics, The University of Tokyo, Tokyo 113-8656, Japan}

\date{\today}

\begin{abstract}

Superpositions of spin helices can yield topological spin textures, such as two-dimensional vortices and skyrmions, 
and three-dimensional hedgehogs. 
Their topological nature and spatial dimensionality depend on the number and relative directions of the constituent helices. 
This allows mutual transformation between the topological spin textures by controlling the spatial anisotropy. 
Here we theoretically study the effect of anisotropy in the magnetic interactions 
for an effective spin model for chiral magnetic metals. 
By variational calculations for both cases with triple and quadruple superpositions, we find that 
the hedgehog lattices, which are stable in the isotropic case, are deformed by the anisotropy, 
and eventually changed into other spin textures with reduced dimension, such as helices and vortices. 
We also clarify the changes of topological properties by tracing the real-space positions of magnetic monopoles 
and antimonopoles as well as the emergent magnetic field generated by the noncoplanar spin textures.
Our results suggest possible control of the topological spin textures, e.g., 
by uniaxial pressure and chemical substitution in chiral materials. 
\end{abstract}


\maketitle

\section{Introduction \label{sec1}}

\begin{figure*}[t]
	\includegraphics[width=\textwidth]{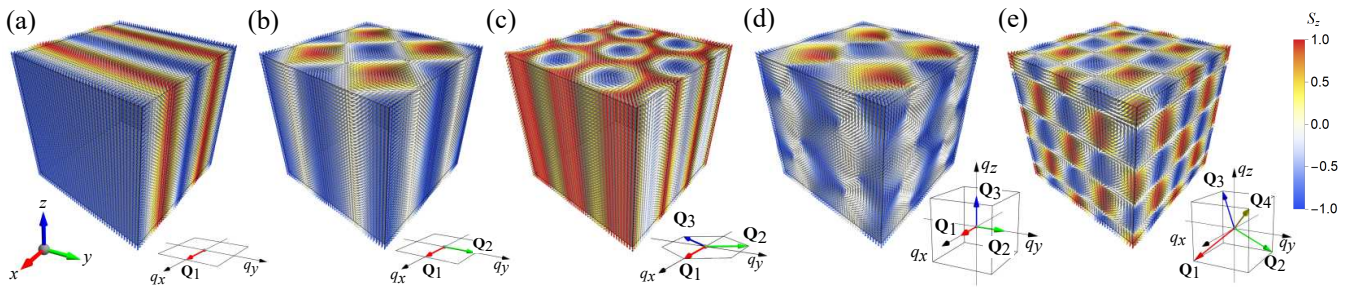}
	\caption{
		\label{fig:fig1}
		Real-space spin configurations of (a) a helical state ($1Q$-H), (b) a vortex crystal composed of two helices ($2Q$-VC), 
		(c) a skyrmion lattice composed of three (SkL), (d) a hedgehog lattice composed of three ($3Q$-HL) and 
		(e) another hedgehog lattice composed of four ($4Q$-HL). 
		The arrows represent the spins and their colors display the $z$ component.
		Each inset shows the wave vectors of the constituent helices.
		}
\end{figure*}

Chiral magnets with noncentrosymmetric crystal structures often show interesting spin textures through the spin-orbit coupling. 
In these systems, the spin-orbit coupling generates an effective antisymmetric exchange interaction, called the Dzyaloshinskii-Moriya (DM) interaction~\cite{Dzyaloshinsky1958,Moriya1960}, which twists the spin configuration.
The simplest example is a helical structure shown in Fig.~\ref{fig:fig1}(a)
~\cite{Yoshimori1959,Moriya1976,Shirane1983,Uchida2006}. 
There are, however, a variety of noncollinear and noncoplanar spin textures 
composed of superpositions of the helices. 
Several examples are shown in Fig.~\ref{fig:fig1}: 
a vortex crystal (VC) [Fig.~\ref{fig:fig1}(b)]~\cite{Hayami2017,Hayami2018,Khanh2020},
a skyrmion lattice (SkL) [Fig.~\ref{fig:fig1}(c)]~\cite{Muhlbauer2009,Yu2010,Tonomura2012}, 
and hedgehog lattices (HLs) [Figs.~\ref{fig:fig1}(d) and \ref{fig:fig1}(e)]
~\cite{Fujishiro2019, Fujishiro2020}. 
The wave vectors of the constituent helices are shown in  each inset.
The VC and SkL have two-dimensional swirling spin textures composed of two and three helices, respectively, whose wave vectors are on the same plane, while the two different types of the HLs have three-dimensional textures composed of 
three and four helices spanning in three dimensions. 
The SkL has been found in the wide range of materials, such as 
B20 compounds (MnSi~\cite{Muhlbauer2009,Tonomura2012}, FeGe~\cite{Yu2011}, and Fe$_{1-x}$Co$_{x}$Si~\cite{Munzer2010,Yu2010}), 
multiferroic insulator Cu$_2$OSeO$_3$~\cite{Seki2012,Adams2012}, and $\beta$-Mn type Co-Zn-Mn alloys~\cite{Tokunaga2015}. 
The HL composed of three helices, which we call 3$Q$-HL, was observed in the B20 compound MnGe~\cite{Tanigaki2015}. 
The other type of HL composed of four helices called 4$Q$-HL was also discovered in MnSi$_{1-x}$Ge$_x$ $(0.3\leq x\leq0.6)$~\cite{Fujishiro2019}.  
Interestingly, the VC was recently observed in a centrosymmetric magnet GdRu$_2$Si$_2$~\cite{Khanh2020}, 
suggesting other stabilization mechanism rather than the DM interaction. 
We also note that the $4Q$-HL was found in the centrosymmetric cubic perovskite SrFeO$_3$~\cite{Ishiwata2020}. 

These spin textures affect the electronic state of the system in several ways. 
One is in the spatial dimensionality inherited from the magnetic textures: 
The helical structure in Fig.~\ref{fig:fig1}(a) has a one-dimensional modulation, the VC in Fig.~\ref{fig:fig1}(b) and 
the SkL in Fig.~\ref{fig:fig1}(c) have two-dimensional ones, and the HLs in Figs.~\ref{fig:fig1}(d) and \ref{fig:fig1}(e) have 
three-dimensional ones. 
The difference in the dimensionality yields the corresponding anisotropy in the electronic state. 
Another way is in the quantum phase of the electron wave function, which is called the Berry phase~\cite{Berry1984}. 
Itinerant electrons coupled with these 
noncollinear and noncoplanar spin textures feel an emergent electromagnetic field through 
the Berry phase, and exhibit unconventional transport and optical responses~\cite{Xiao2010,Nagaosa2013}.
In addition, some of the spin textures have topologically nontrivial structures, 
which are robust against continuous deformation and characterized by topological numbers. 
For instance, the SkL is characterized by the so-called skyrmion number, and the HLs are by the monopole charges~\cite{Volovik1987,Kanazawa2016,Okumura2020tracing}. 
In particular, the HLs are regarded as a periodic array of hedgehog- and antihedgehog-like spin textures 
that are sources and sinks of the emergent magnetic field (monopoles and antimonopoles), respectively. 
Such topological nature is also relevant to the electronic properties~\cite{Takashima2014,Kanazawa2016,Zhang2016}.

The question that we address in this study is how to control the spin textures with 
different dimensionality and topology. 
Once one can switch them from one to another, it may be able to manipulate not only the magnetism  
but also the electronic state, transport, and optical responses. 
An example of such a control was found in a certain class of chiral magnets, e.g., the B20 compounds, by applying a magnetic field; 
the one-dimensional helical state at zero field is changed into a two-dimensional SkL 
in an applied field~\cite{Muhlbauer2009, Yu2010, Tonomura2012}, 
and correspondingly, the system exhibits an unconventional Hall effect, which is called the topological Hall effect, 
due to the effective magnetic field arising from the topological noncoplanar spin texture in the SkL. 
Similar switching was achieved by spatial geometry and anisotropy; 
a conical state in a bulk sample is taken over by the SkL in thin films~\cite{Yu2011,Tonomura2012,Yu2015}. 
In addition, interesting transitions among helical, SkL, and HLs were discovered in MnSi$_{1-x}$Ge$_x$ by chemical substitution~\cite{Fujishiro2019}. 
Such interesting possibilities to control the dimensionality and topology, however, have not been systematically investigated thus far, especially including the three-dimensional HLs. 

In this paper, we theoretically study such a control of the spin textures 
focusing on the effect of spatial anisotropy.
For an effective spin model with long-range interactions stemming from 
itinerant nature of electrons~\cite{Okumura2020}, we clarify the ground-state phase diagram by variational calculations 
while changing the anisotropy in the interactions. 
We find that the HLs, which are stable in the isotropic case, change their magnetic textures with dimensional reduction 
by the anisotropy.  
In the $3Q$-HL case, the three-dimensional magnetic texture turns into a two-dimensional VC 
for the hard-axis anisotropy along the helix direction, while into another type of VC and 
eventually a one-dimensional helical state for the easy-axis anisotropy. 
Meanwhile, the $4Q$-HL changes into the $3Q$-HL and a helical state for the hard- and easy-axis anisotropy, respectively. 
Tracing the spatial positions of the hedgehogs and antihedgehogs, 
we find several types of topological transitions with their creation and annihilation while changing the anisotropy.
We also clarify how the flow of the emergent magnetic field is modulated through the phase transitions. 
Our results demonstrate that the spatial anisotropy is useful to control not only the magnetic structures including their dimensionality 
but also electric, transport, and optical properties arising from the topological nature through the emergent magnetic field. 

The rest of the paper is organized as follows.
In Sec.~\ref{sec:sec2}, we introduce the effective spin model with spatial anisotropy in the long-range interactions. 
In Sec.~\ref{sec:sec3}, we describe the variational method for investigating the ground state of the model.
In Sec.~\ref{sec:sec4}, we show the results for the phase diagram in a wide parameter space (Sec.~\ref{sec:sec4-1}), 
the phase transitions driven by the anisotropy (Sec.~\ref{sec:sec4-2}), and corresponding changes of 
the topological properties associated with the monopoles and antimonopoles as well as the emergent magnetic fields 
(Sec.~\ref{sec:sec4-3}). 
Section~\ref{sec:sec5} is devoted to the summary.

\section{MODEL \label{sec:sec2}}

We consider an effective spin model for itinerant chiral magnets that includes long-range interactions arising from the spin-charge coupling and the spin-orbit coupling. 
The model was originally derived from the perturbation expansion for the Kondo lattice model with 
an antisymmetric spin-orbit coupling~\cite{Hayami2017,Okumura2020}. 
The previous studies showed that this model exhibits  
VC and SkL in two dimensions~\cite{Ozawa2017,Hayami2017,Hayami2018} and HLs in three dimensions~\cite{Okumura2020}. 
The Hamiltonian reads
\begin{eqnarray}
	\mathcal{H}=
	&2 &\sum_{\eta} \left[
	-J_{\eta} {\bf S}_{{\bf Q}_{\eta}} \cdot {\bf S}_{-{\bf Q}_{\eta}}
	+\frac{K_{\eta}}{N}\left({\bf S}_{{\bf Q}_{\eta}}\cdot{\bf S}_{-{\bf Q}_{\eta}}\right)^2 \right. \notag \\
	&& \qquad \left. +i{\bf D}_{\eta}\cdot\left({\bf S}_{{\bf Q}_{\eta}}\times{\bf S}_{-{\bf Q}_{\eta}}\right) \right].
	\label{eq:JKmodel}
\end{eqnarray}
All the interactions are defined in terms of the Fourier component of spins, 
${\bf S}_{\bf q}=\frac{1}{\sqrt{N}}\sum_{l}\ {\bf S}_{{\bf r}_l} e^{-i{\bf q}\cdot{\bf r}_l}$, 
where ${\bf S}_{{\bf r}_l}$ 
denotes the spin at a real-space position ${\bf r}_l$ and  $N$ is the number of spins in the system. 
The sum in Eq.~(\ref{eq:JKmodel}) is taken for a particular set of the wave vectors ${\bf Q}_{\eta}$ derived from the electronic band structure 
in the original itinerant electron model (see below). 
We consider the model on a cubic lattice, following the previous study~\cite{Okumura2020}.

The first term in Eq.~(\ref{eq:JKmodel}) represents the Ruderman-Kittel-Kasuya-Yosida (RKKY) interaction 
derived from the second-order perturbation with respect to the spin-charge coupling; $J_{\eta}>0$.
This prefers a helical magnetic order with wave vector ${\bf Q}_{\eta}$.
The second term is the biquadratic interaction obtained from the fourth-order perturbation~\cite{Akagi2012,Hayami2017}. 
The coupling constant $K_\eta$ is positive, which favors a multiple-$Q$ state composed of a superposition of the helices 
with different ${\bf Q}_{\eta}$, such as VC, SkL, and HLs. 
The third term is a DM-type interaction originating from the second-order perturbation in terms of both spin-charge and 
antisymmetric spin-orbit couplings.
Following the previous studies~\cite{Hayami2017,Okumura2020}, we set the directions of 
${\bf D}_{\eta}$ as ${\bf D}_{\eta}\parallel {\bf Q}_{\eta}$, namely, ${\bf D}_\eta = D_\eta {\bf Q}_\eta/|{\bf Q}_\eta|$.
This term prefers a proper screw spin texture propagating in the direction of ${\bf Q}_{\eta}$ 
with chirality depending on the sign of ${\bf D}_{\eta}\cdot{\bf Q}_{\eta}$.

In this Hamiltonian, all the interactions are specified by 
the wave vectors ${\bf Q}_{\eta}$, which are set by multiple maxima in the spin-dependent bare susceptibility 
of itinerant electrons in the perturbation theory~\cite{Hayami2017}. 
Following the previous study~\cite{Okumura2020}, we assume two types of ${\bf Q}_{\eta}$: 
One is a set of cubic wave vectors which are orthogonal to each other as
${\bf Q}_1=(Q,0,0)$, ${\bf Q}_2=(0,Q,0)$, and ${\bf Q}_3=(0,0,Q)$ [see the inset of Fig.~\ref{fig:fig1}(d)], and 
the other is a set of tetrahedral wave vectors as
${\bf Q}_1=(Q,-Q,-Q)$, ${\bf Q}_2=(-Q,Q,-Q)$, ${\bf Q}_3=(-Q,-Q,Q)$, and ${\bf Q}_4=(Q,Q,Q)$ [see the inset of Fig.~\ref{fig:fig1}(e)].
In the following analysis, we set $Q=\pi/12$ without loss of generality.

To investigate the effects of spatial anisotropy, we introduce the anisotropy in the 
coupling constants $J_\eta$, $K_\eta$, and $D_{\eta}$ in Eq.~(\ref{eq:JKmodel}).
For the 3$Q$ case, we introduce the anisotropy along the [001] direction parallel to ${\bf Q}_3$ as 
$J_{\eta}=J$,  $D_{\eta}=D$, and $K_{\eta}=K$ for $\eta=1$ and $2$, 
and $J_\eta=J'$, $D_\eta=D'$, and $K_\eta=K'$ for $\eta=3$. 
Meanwhile, for the 4$Q$ case, we introduce the anisotropy along the [111] direction parallel to ${\bf Q}_4$ as 
$J_{\eta}=J$, $D_{\eta}=D$, and $K_{\eta}=K$ for $\eta=1,2,3$, 
and $J_\eta=J'$, $D_\eta=D'$, and $K_\eta=K'$ for $\eta=4$. 
In both cases, for simplicity, we assume the relations $D'/D=J'/J$ and $K'/K=(J'/J)^2$ 
considering the order of the perturbation expansion. 
Thus, $J'/J>1$ ($<1$) corresponds to the easy-axis (hard-axis) anisotropy along the [001] and [111] directions 
for the $3Q$ and $4Q$ cases, respectively. 

\section{METHOD \label{sec:sec3}}

We study the ground state of the model in Eq.~(\ref{eq:JKmodel}) by a variational method. 
The method is an extension of that used in the previous study~\cite{Okumura2020} to the anisotropic cases. 
We consider the following three types of spin textures as the variational states.
The first one is a superposition of sinusoidal waves given by
\begin{eqnarray}
	{\bf S}_{{\bf r}_l}\propto \sum_{\eta=1}^{n}
	a_{\eta}{\bf  e}_{\eta}^0 \cos \mathcal{Q}_{\eta l},
	\label{eq:nonchiralstate}
\end{eqnarray}
where ${\bf e}_{\eta}^0$ is the unit vector parallel to ${\bf Q}_{\eta}$ and 
$\mathcal{Q}_{\eta l}={\bf Q}_{\eta}\cdot{\bf r}_{l}+\phi_{\eta}$;
$n=3$ and $4$ for the $3Q$ and $4Q$ cases, respectively. 
$a_{\eta} $ and $\phi_{\eta}$ denote the amplitude and phase of the 
${\bf Q}_{\eta}$ component, respectively. 
As this is the nonchiral state which has no energy gain from the DM-type interaction, we call it the 
$mQ$ nonchiral state ($mQ$-NC), where $m =1,2,\cdots, n$ is the number of the components with nonzero $a_\eta$ in Eq.~(\ref{eq:nonchiralstate}). 
The second variational state is given by
\begin{eqnarray}
	{\bf S}_{{\bf r}_l}\propto \sum_{\eta=1}^{n}
	a_{\eta}\left({\bf  e}_{\eta}^1 \cos \mathcal{Q}_{\eta l}+{\bf  e}_{\eta}^2 \sin \mathcal{Q}_{\eta l}\right),
	\label{eq:chiralstate}
\end{eqnarray}
where ${\bf e}_{\eta}^1$ and ${\bf e}_{\eta}^2$ are the unit vectors which satisfy 
${\bf e}_{\eta}^{1}\times{\bf e}_{\eta}^{2}={\bf e}_{\eta}^{0}$; $n=3$ and $4$ for the $3Q$ and $4Q$ cases, respectively, as in Eq.~(\ref{eq:nonchiralstate}).
Here, we set 
${\bf e}_1^1=\hat{{\bf y}},$ 
${\bf e}_2^1=\hat{{\bf z}},$ 
and 
${\bf e}_3^1=\hat{{\bf x}}$ 
for the 3$Q$ case, and 
${\bf e}_{\eta}^1 = \hat{{\bf z}}\times{\bf e}_{\eta}^0/|\hat{{\bf z}}\times{\bf e}_{\eta}^0|$ 
for the 4$Q$ case, 
where $\hat{{\bf x}}$, $\hat{{\bf y}}$, and $\hat{{\bf z}}$ are the unit vectors in the $xyz$ coordinates.
This is a chiral multiple-$Q$ state, which has an energy gain from the DM-type interaction 
in contrast to the nonchiral one in Eq.~(\ref{eq:nonchiralstate}).
We call it the 1$Q$ helical state (1$Q$-H) when only one of $a_\eta$ is nonzero, while
the 2$Q$ vortex crystal (2$Q$-VC) when two of them are nonzero.
Meanwhile, for the states with more than two nonzero components, we have topologically different states depending on 
the existence of hedgehogs and antihedgehogs (see below); 
we call the states with hedgehogs and antihedgehogs the 3$Q$ and 4$Q$ hedgehog lattices ($3Q$- and 4$Q$-HLs),
while otherwise simply the 3$Q$ and 4$Q$ states. 
The third variational state is called the double-$Q$ chiral stripe (2$Q$-CS)
found in the previous study~\cite{Ozawa2016}, 
which is given by 
\begin{eqnarray}
	{\bf S}_{{\bf r}_l}\propto 
	\sqrt{1-u_{\eta' l}^2}\left({\bf  e}_{\eta}^1 \cos \mathcal{Q}_{\eta l}+{\bf  e}_{\eta}^2 \sin \mathcal{Q}_{\eta l}\right)
	+u_{\eta' l} {\bf e}_{\eta}^0,
	\label{eq:chiralstripe}
\end{eqnarray}
where $u_{\eta' l} = v\sin\mathcal{Q}_{\eta' l}$; 
$v$ denotes the amplitude of the sinusoidal wave, and
$\eta$ and $\eta'$ take two of three and four components 
in the $3Q$ and $4Q$ cases, respectively. 

In the variational calculations, $a_{\eta}$, $\phi_{\eta}$, and $v$ are treated as variational parameters 
that are optimized to find the lowest-energy state.  
$a_{\eta}$ and $v$ are varied from 0 to 1, while  $\phi_{\eta}$ from $0$ to 
$Q$ for the 3$Q$ case and $0$ to $2\pi$ for the 4$Q$ case. 
We prepare each variational state with the normalization of the spin length at every site as $|{\bf S}_{{\bf r}_l}|=1$, optimize the energy with respect to the variational parameters, and then compare the energies among different variational states. 
Note that the optimized energy is unchanged by exchanging $\eta$ and $\eta'$, except for the $2Q$-CS state 
including the helix running along the anisotropic axis in $\eta$ or $\eta'$.
In the following analysis, we impose the following constraints on the variational parameters $a_{\eta}$ 
from the symmetry point of view: 
$a_1=a_2$ for all the multiple-$Q$ states in the $3Q$ case, 
and $a_1=a_2$ (or equivalently, $a_2=a_3$ or $a_3=a_1$) for the 2$Q$-NC and VC in the $4Q$ case.
In addition, we set $a_1=a_2=a_3$ for all the $3Q$ and $4Q$ states including the HLs for the $4Q$ case. 
In all cases, $\sum_{\eta} a_{\eta}^2 = 1$.
We perform the variational calculations on the cubic lattice with $N=24^3$ sites under the periodic boundary conditions in all three directions. 
Note that there is no finite-size effects as $Q=\pi/12$ induces magnetic orders with period of 24.

For the optimized states, we calculate the magnetic moment with wave vector ${\bf Q}_\eta$~\footnote{
We adopt a slightly different definition for $m_{{\bf Q}_{\eta}}$ from that in Ref.~\cite{Okumura2020}; 
we take a symmetric summation with respect to $\pm{\bf Q}_{\eta}$, 
which results in just a difference of factor $\sqrt{2}$.},
\begin{eqnarray}
	m_{{\bf Q}_{\eta}}=\sqrt{\frac{S({\bf Q}_{\eta})+S(-{\bf Q}_{\eta})}{N}},
	\label{eq:m_q}
\end{eqnarray}
where $S({\bf q})$ is the spin structure factor defined by
\begin{eqnarray}
	S({\bf q})=\frac{1}{N}\sum_{l,l'} {\bf S}_{{\bf r}_l}\cdot{\bf S}_{{\bf r}_{l'}}e^{-i{\bf q}\cdot({\bf r}_l-{\bf r}_{l'})}.
\end{eqnarray}

We also calculate a topological quantity called the monopole charge, which is defined 
for each unit cube composed of the neighboring eight sites on the cubic lattice as~\cite{Okumura2020tracing,Okumura2020}
\begin{eqnarray}
	Q_{\rm m}({\bf R}_j)=\frac{1}{4\pi}\sum_{p \, \in \, j{\rm th}\, {\rm unit}\, {\rm cube}}
	\Omega_p, \label{eq:monopole_charge}
\end{eqnarray}
where ${\bf R}_j$ is the center position of $j$th unit cube and $\Omega_p$ is the solid angle spanned by four spins
on the $p$th square plaquette surrounding the unit cube $(p=1,2,\ldots, 6)$.
$\Omega_{p}$ is calculated by dividing the plaquette into two triangles $\triangle_i\ (i=1,2)$ and taking the sum of
the solid angles of three spins in the triangles as $\Omega_p= \Omega_{\triangle_1} + \Omega_{\triangle_2}$, where 
\begin{eqnarray}
	\Omega_{\triangle_i}=2\tan^{-1}
	\left[
	\frac{{\bf S}_1\cdot({\bf S}_2\times{\bf S}_3)}
	{1+{\bf S}_1\cdot{\bf S}_2+{\bf S}_2\cdot{\bf S}_3+{\bf S}_3\cdot{\bf S}_1}
	\right]. 
\label{eq:solid_angle}
\end{eqnarray}
Here, ${\bf S}_1$, ${\bf S}_2$, and ${\bf S}_3$ are three spins on the triangle $\triangle_i$ in the clockwise order 
viewed from the center of the cube. 
The argument in $\tan^{-1}$ is taken in the range of $[-\pi,\pi )$.
The monopole charge in Eq.~(\ref{eq:monopole_charge}) signals the position of hedgehogs and antihedgehogs 
by $Q_{\rm m}({\bf R}_j)=+1$ and $-1$, respectively. 
We also compute the total number of hedgehogs and antihedgehogs in the magnetic unit cell by
\begin{eqnarray}
	N_{\rm m}=\sum_{j \, \in \, {\rm magnetic} \, {\rm unit} \, {\rm cell}}\ |Q_{\rm m}({\bf R}_j)|.
\end{eqnarray}

We also calculate the scalar spin chirality which gives the emergent magnetic field through the Berry phase.
We compute ${\boldsymbol \chi}_{{\bf r}_l} = (\chi_{{\bf r}_l}^x,\chi_{{\bf r}_l}^y,\chi_{{\bf r}_l}^z)$ at each site, whose component 
is defined by the summation of the spin triple products on the adjacent four triangles as~\cite{Okumura2020tracing,Okumura2020}
\begin{align}
	\chi_{{\bf r}_l}^{\alpha} = 
		\frac{1}{8}\sum_{\beta \gamma \nu_{\beta} \nu_{\gamma}}
		\varepsilon^{\alpha\beta\gamma}\nu_{\beta}\nu_{\gamma}
		{\bf S}_{{\bf r}_l}\cdot( {\bf S}_{{\bf r}_l+\nu_{\beta}{\bf e}_{\beta}}
		\times {\bf S}_{{\bf r}_l+\nu_{\gamma}{\bf e}_{\gamma}} ).
\label{eq:chi}
\end{align}
Here, $\beta$ and $\gamma$ are the directions perpendicular to $\alpha$, and 
$\nu_{\beta(\gamma)}=\pm1$. 

The definitions in Eqs.~(\ref{eq:monopole_charge}), (\ref{eq:solid_angle}), and (\ref{eq:chi}) indicate that 
the hedgehogs and antihedgehogs are regarded as magnetic monopoles and antimonopoles 
that are the sources and sinks of the emergent magnetic field, respectively. 
Real-space mapping of the positions of the monopoles and antimonopoles with 
${\boldsymbol \chi}_{{\bf r}_l}$ manifests the flow of the emergent magnetic field; see Sec.~\ref{sec:sec4-3}.

\section{RESULTS \label{sec:sec4}}

In this section, we show the results obtained by the variational calculations for the model in Eq.~(\ref{eq:JKmodel}). 
In Sec.~\ref{sec:sec4-1}, we show the phase diagrams on the $D$($D'$)-$K$($K'$) plane while changing the anisotropy. 
Next, in Sec.~\ref{sec:sec4-2}, we show how the magnetic and topological properties change 
through the phase transitions caused by changing the anisotropy. 
Finally, in Sec.~\ref{sec:sec4-3}, we show real-space distributions of hedgehogs and antihedgehogs, and 
the flow of the scalar spin chirality ${\boldsymbol \chi}_{{\bf r}_l}$. 

\subsection{Phase diagram \label{sec:sec4-1}}
\begin{figure}[t]
	\includegraphics[width=1.0\columnwidth]{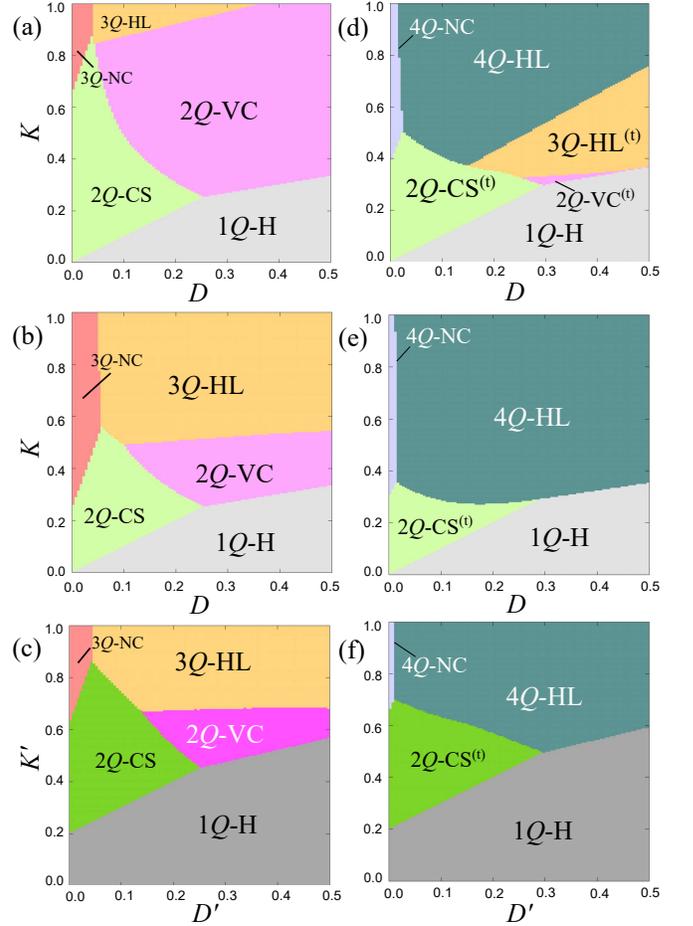}
	\caption{
		\label{fig:fig2}
		Phase diagrams of the model in Eq.~(\ref{eq:JKmodel}) obtained  
		by the variational calculations 
		for (a)-(c) the $3Q$ case and (d)-(f) the $4Q$ case:
		(a),(d) $J'/J=0.8$ (hard-axis), (b),(e) $J'/J=1.0$ (isotropic), and (c),(f) $J/J'=0.8$ (easy-axis). 
		 $1Q$-H, $2Q$-VC, $2Q$-CS, $3Q(4Q)$-HL, and $3Q(4Q)$-NC represent 
		the $1Q$ helical, the $2Q$ vortex crystal, the $2Q$ chiral stripe,
		the $3Q(4Q)$ hedgehog lattice, and the  
		$3Q$($4Q$) nonchiral states, respectively.
		The light and dark colors for $2Q$-VC and $1Q$-H denote the differences in the constituent wave vectors.
		}
\end{figure}

First, we show the phase diagrams obtained by the variational calculations for the 3$Q$ case in 
Figs.~\ref{fig:fig2}(a)-\ref{fig:fig2}(c) and for the 4$Q$ case in Figs.~\ref{fig:fig2}(d)-\ref{fig:fig2}(f): 
(a) and (d) are for the hard-axis anisotropy with $J'/J=0.8$, 
(b) and (e) are for the isotropic cases with $J'/J=1.0$, and 
(c) and (f) are for the easy-axis anisotropy with $J/J'=0.8$. 
We plot the results for $D$ and $K$ when $J'/J\leq1.0$, while for $D'$ and $K'$ when $J'/J>1.0$, 
with the relations $D'/D=J'/J$ and $K'/K=(J'/J)^2$. 
Note that the $2Q$-NC, VC, CS, and $3Q$-HL are different states between 
the $3Q$ and $4Q$ cases due to the different sets of ${\bf Q}_\eta$.

The results for the isotropic cases in Figs.~\ref{fig:fig2}(b) and \ref{fig:fig2}(e) agree with those in the previous study~\cite{Okumura2020}; 
the 3$Q$- and 4$Q$-HLs appear in a wide range of $D$ and $K$, indicating that both spin-orbit 
and spin-charge couplings play an important role in the stabilization of the HLs. 
When we introduce the hard-axis anisotropy, as shown in Figs.~\ref{fig:fig2}(a) and \ref{fig:fig2}(d), 
the regions of these HLs are shrunk, and instead, the $2Q$-VC and CS (the $2Q$-CS and the $3Q$-HL) 
are extended for the $3Q$ ($4Q$) case; in the $3Q$ case, the $2Q$-VC and CS states are composed of 
${\bf Q}_1$ and ${\bf Q}_2$, while in the $4Q$ case, the $2Q$-CS is composed of 
two out of ${\bf Q}_1$, ${\bf Q}_2$, and ${\bf Q}_3$, and the $3Q$-HL is of ${\bf Q}_1$, ${\bf Q}_2$, and ${\bf Q}_3$.
On the other hand, when we introduce the easy-axis anisotropy, as shown in Figs.~\ref{fig:fig2}(c) and \ref{fig:fig2}(f), 
the $1Q$-H is extended to the regions of the HLs, which is composed of ${\bf Q}_3$ (${\bf Q}_4$) in ths 3$Q$ (4$Q$) case.
These results are understood from the fact that the hard(easy)-axis anisotropy tends to destabilize (stabilize) 
the helix propagating along ${\bf Q}_3 \parallel [001]$ and ${\bf Q}_4 \parallel [111]$ for the $3Q$ and $4Q$ cases, respectively. 

\subsection{Phase transitions driven by anisotropy \label{sec:sec4-2}}

\begin{figure}[tb]
	\includegraphics[width=0.98\columnwidth]{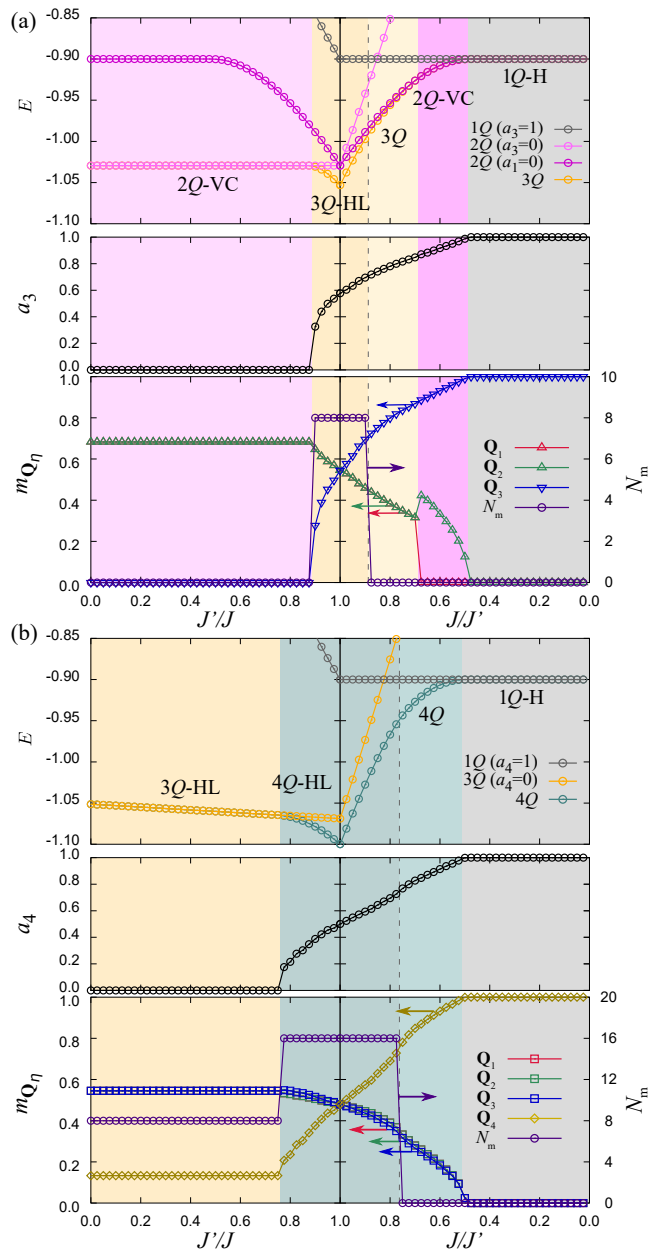}%
	\caption{
		\label{fig:fig3}
		Anisotropy dependences of the variational energy per site, $E$, 
		the amplitude of the helix in the $J'$ direction, $a_3$ or $a_4$, 
		the magnetic moment with the wave vector ${\bf Q}_{\eta}$, $m_{{\bf Q}_{\eta}}$, and
		the number of hedgehog and antihedgehog, $N_{\rm m}$, for the (a)  
		$3Q$ and (b) $4Q$ cases. 
		The dashed lines represent the topological transitions with the changes of $N_{\rm m}$.
		The results are calculated for $D=0.3$ and $K=0.8$ in Eq.~(\ref{eq:JKmodel}). 
		We take $J$($J'$) as the energy unit for $J'/J<1$($>1$).
		}
\end{figure}

The results in Fig.~\ref{fig:fig2} indicate that the system exhibits phase transitions from the isotropic HLs to 
other magnetic states by introducing the anisotropy. 
We discuss such phase transitions in detail, for $D=0.3$ and $K=0.8$ where the $3Q$- and $4Q$-HLs 
are stable in the isotropic $3Q$ and $4Q$ cases, respectively. 
Figures~\ref{fig:fig3}(a) and \ref{fig:fig3}(b) show the results for the $3Q$ and $4Q$ cases, respectively:  
anisotropy dependences of the optimized energy per site, $E$, 
the variational parameters $a_3$ and $a_4$, the magnetic moments $m_{{\bf Q}_{\eta}}$, 
and the total number of hedgehogs and antihedgehogs in the magnetic unit cell, $N_{\rm m}$. 

Let us first discuss the $3Q$ case in Fig.~\ref{fig:fig3}(a). 
In this case, we identify five different phases from the energy comparison and the topological nature. 
The energy comparison is plotted in the upper panel of Fig.~\ref{fig:fig3}(a) for the $1Q$-H state with $a_3=1$ (gray), 
$2Q$-VC with $a_3=0$ (light pink), $2Q$-VC with $a_1=0$ (or $a_2=0$) (dark pink), and $3Q$ (orange). 
The values of $a_3$ and $m_{{\bf Q}_\eta}$ for the lowest-energy state are shown in the middle and lower panels. 
In the region where the $3Q$ variational state has the lowest energy, we distinguish two topologically different phases 
by $N_{\rm m}$ plotted in the lower panel. We describe the details below. 

In the isotropic case, the system exhibits the $3Q$-HL with the same amplitude as indicated by 
$a_3 = 1/\sqrt{3}$ ($=a_1=a_2$) and $m_{{\bf Q}_1} = m_{{\bf Q}_2} = m_{{\bf Q}_3}$ 
as plotted in the middle and lower panels of Fig.~\ref{fig:fig3}(a), respectively. 
This state is topologically nontrivial as it includes four pairs of hedgehogs and antihedgehogs in the magnetic unit cell, 
i.e., $N_{\rm m}=8$, as plotted in the lower panel of Fig.~\ref{fig:fig3}(a). 
We find that the phases of the three ordering vectors are $\phi_1=\phi_2=\phi_3=Q/2$, which indicates that 
the hedgehogs and antihedgehogs are located at the centers of the cubes so as to avoid the lattice sites~\cite{Okumura2020}. 

When we introduce the hard-axis anisotropy by taking $J'/J<1$, the $3Q$-HL remains stable down to 
$J'/J \simeq 0.8875$ while the spin texture is modulated by reflecting the anisotropy as 
$a_3<1/\sqrt{3}$ ($a_1=a_2>1/\sqrt{3}$) and $m_{{\bf Q}_1}=m_{{\bf Q}_2}>m_{{\bf Q}_3}$. 
In the anisotropic HL, $\phi_{\eta}$ takes $0$ or $Q/2$, which is chosen so as to locate the hedgehogs and antihedgehogs at the interstitial positions appropriately. 
For $J'/J \lesssim 0.8875$, the $3Q$-HL is taken over by the 2$Q$-VC with $a_3=0$ ($a_1=a_2=1/\sqrt{2}$) 
and $m_{{\bf Q}_3} = 0$, as shown in the left hand side of Fig.~\ref{fig:fig3}(a). 
In this state, both $\phi_1$ and $\phi_2$ are optimized to be $Q/2$ so that the vortex and antivortex cores are located 
at the centers of the square plaquettes.  
At this phase transition, $N_{\rm m}$ changes from $8$ to $0$, as shown in the lower panel of Fig.~\ref{fig:fig3}(a). 
Thus, the results indicate that the transition occurs from the topologically nontrivial $3Q$-HL to the topologically 
trivial $2Q$-VC by the hard-axis anisotropy. 

On the other hand, when we introduce the easy-axis anisotropy as $J'/J>1$, the system shows more complicated 
multiple phase transitions, as shown in the right hand side of Fig.~\ref{fig:fig3}(a). 
Note that the right hand side is plotted for $J/J'$ instead of $J'/J$ to cover all the parameter range to the limit of $J'/J \to \infty$. 
The 3$Q$-HL with $N_{\rm m}=8$ is stable up to $J/J' \simeq 0.8875$ while the spin configuration is modulated 
as $a_3 > a_1=a_2$ and $m_{{\bf Q}_3}>m_{{\bf Q}_1}=m_{{\bf Q}_2}$. 
At $J/J' \simeq 0.8875$, the topologically nontrivial 3$Q$-HL changes into the topologically trivial $3Q$ state with $N_{\rm m}=0$. 
We note that $\phi_\eta$ takes $0$ or $Q/2$ for $J/J'\gtrsim 0.7875$ similar to those for $0.8875\lesssim J'/J<1$, whereas 
$\phi_1=\phi_2=\phi_3=0$ for $J/J'\lesssim 0.7875$; the origin of the phase shifts is unclear, while in both regions, the hedgehogs and antihedgehogs 
are located at the interstitial positions.
With a further decrease of $J/J'$ to $\simeq 0.6875$, the 3$Q$ state is taken over by the 2$Q$-VC with 
$a_1=0$ and $a_3>a_2>0$, and $m_{{\bf Q}_1}=0$ and $m_{{\bf Q}_3}>m_{{\bf Q}_2}>0$, as shown in the middle and lower panels of Fig.~\ref{fig:fig3}(a). 
This state is different from the 2$Q$-VC stable for $J'/J \lesssim 0.8875$; it has no singular points where the spin length vanishes 
since it is composed of two helices with different amplitude. 
We note that $\phi_{\eta}$ varies with $J'/J$ in this state.
Finally, for $J/J' \lesssim 0.4875$, the 2$Q$-VC is replaced by the 1$Q$-H with $a_3=1$ and $a_1=a_2=0$, and 
$m_{{\bf Q}_3}=1$ and $m_{{\bf Q}_1}=m_{{\bf Q}_2}=0$, which is stable to the easy-axis limit, $J/J' \to 0$. 

Next, we discuss the 4$Q$ case in Fig.~\ref{fig:fig3}(b).
In this case, we identify four phases. The energy comparison is plotted in the top panel of Fig.~\ref{fig:fig3}(b) for $1Q$-H with $a_4=1$ (gray), $3Q$ with $a_4=0$ (orange), 
and $4Q$ (green). As in the $3Q$ case above, we obtain two topologically different phases in the $4Q$ region from the values of $N_{\rm m}$ plotted 
in the lower panel of Fig.~\ref{fig:fig3}(b), as described below.

In the isotropic case, the system exhibits the 4$Q$-HL with equal amplitude of the four helices, i.e., 
$a_4=1/2$ ($=a_1=a_2=a_3$) and $m_{{\bf Q}_1}=m_{{\bf Q}_2}=m_{{\bf Q}_3}=m_{{\bf Q}_4}$. 
This state is topologically nontrivial as it includes eight pairs of hedgehogs and antihedgehogs, 
i.e., $N_{\rm m}=16$, as shown in the lower panel of Fig.~\ref{fig:fig3}(b). 
We find that $\phi_\eta$ satisfy the relation $\sum_{\eta=1}^{4}\phi_{\eta} = (2k+1)\pi$ ($k$ is an integer) to place the hedgehogs and antihedgehogs at the centers of the cubes~\cite{Okumura2020}. 
When we introduce the hard-axis anisotropy by taking $J'/J<1$,  
the 4$Q$-HL remains stable down to $J'/J \simeq 0.7625$, while the spin texture is modulated 
as $a_4<1/2$ ($a_1=a_2=a_3>1/2$) and 
$0 < m_{{\bf Q}_4} < m_{{\bf Q}_1} = m_{{\bf Q}_2} = m_{{\bf Q}_3} $. 
For $J'/J \lesssim 0.7625$, the 4$Q$-HL is taken over by the 3$Q$-HL 
with $a_1=a_2=a_3=1/\sqrt{3},~a_4=0$, and $m_{{\bf Q}_1} = m_{{\bf Q}_2} = m_{{\bf Q}_3} > m_{{\bf Q}_4}$, 
as shown in the left hand side of Fig.~\ref{fig:fig3}(b).
The nonzero small value of $m_{{\bf Q}_4}$ despite $a_4=0$ is a consequence of the normalization of the spin length (see Sec.~\ref{sec:sec3}).
In this state, $\phi_{\eta}$ are optimized as $\phi_1=\phi_2=\phi_3=Q/2$.
At this phase transition, $N_{\mathrm m}$ changes from $16$ to $8$.
Note that the magnetic unit cell of the 3$Q$-HL becomes four times smaller than that of the $4Q$-HL; 
it is rhombohedral since the three ordering wave vectors, $ {\bf Q}_1$, ${\bf Q}_2$, and ${\bf Q}_3 $, 
are not orthogonal to each other.

On the other hand, when we introduce the easy-axis anisotropy by $J'/J>1$, 
the system shows two phase transitions as shown in the right hand side of Fig.~\ref{fig:fig3}(b).
The 4$Q$-HL is stable up to $J/J' \simeq 0.7625$, while the spin configuration is modulated 
reflecting the easy-axis anisotropy as $a_1=a_2=a_3<1/2<a_4$, and $m_{{\bf Q}_1} = m_{{\bf Q}_2} = m_{{\bf Q}_3} < m_{{\bf Q}_{4}}$.
At $J/J' \simeq 0.7625$, the topologically nontrivial $4Q$-HL with $N_{\rm m}=16$ is replaced by 
the topologically trivial $4Q$ state with $N_{\rm m}=0$, as shown in the lower panel of Fig.~\ref{fig:fig3}(b).
In these anisotropic 4$Q$-HL and 4$Q$ states, the values of $\phi_{\eta}$ depend on $J'/J$, but the sum always satisfies 
$\sum_{\eta=1}^{4}\phi_{\eta} = (2k+1)\pi$, as for the isotropic $4Q$-HL. 
While further decreasing $J/J'$, the 4$Q$ state is taken over by the 1$Q$-H with $a_1=a_2=a_3=0$ and $a_4=1$, 
and $m_{{\bf Q}_1} = m_{{\bf Q}_2} = m_{{\bf Q}_3} =0$ and $m_{{\bf Q}_4} = 1$ for $J/J'\lesssim 0.5125$.

\subsection{Topological properties \label{sec:sec4-3}}

\begin{figure*}[t]
	\includegraphics[width=2.0\columnwidth]{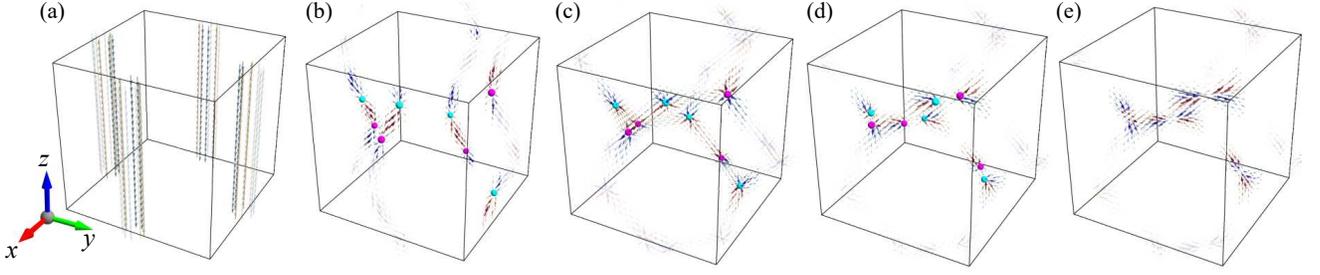}%
	\caption{
	\label{fig:fig4}
	Real-space configurations of the scalar spin chirality ${\boldsymbol \chi}_{{\bf r}_l}$ (arrows), and 
	the monopoles and antimonopoles (magenta and cyan balls, respectively) in a magnetic unit cell 
	obtained from the variational calculations for the $3Q$ case:
	(a) $J'/J=0.875$, (b) $J'/J=0.9$, (c) $J'/J=1.0$, (d) $J/J'=0.9$, and (e) $J/J'=0.875$.
	${\boldsymbol \chi}_{{\bf r}_l}$ are plotted when their lengths are longer than a certain threshold for better visibility. 
	The color of the arrows represents the $z$ component of ${\boldsymbol \chi}_{{\bf r}_l}$; 
	pale color is used for the arrows in the neighboring magnetic unit cells.
	}
\end{figure*}

\begin{figure*}[t]
	\includegraphics[width=2.0\columnwidth]{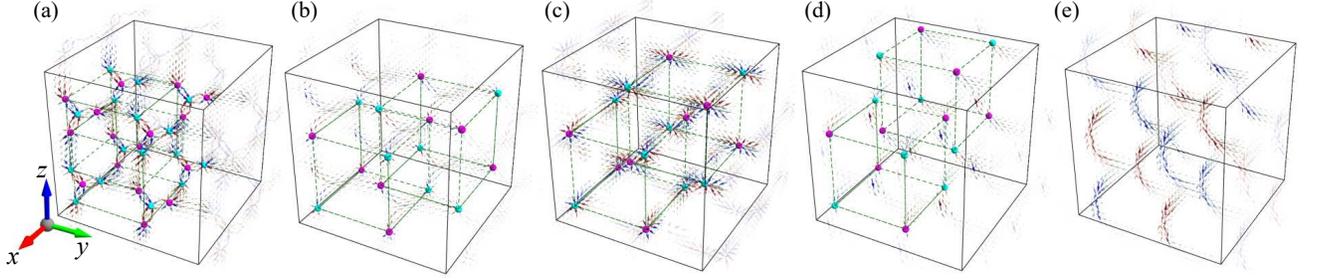}%
	\caption{
	\label{fig:fig5}
	Similar plots for the $4Q$ case: (a) $J'/J=0.75$, (b) $J'/J=0.775$, (c) $J'/J=1.0$, (d) $J/J'=0.775$, and (e) $J/J'=0.75$.
	The notations are common to those in Fig.~\ref{fig:fig4}. 
	Note that the magnetic unit cell is one fourth of the whole cube in (a). 
	The green dashed cubes are the guides for eyes; see the text for details. 
	}
\end{figure*}

In the previous section, we found multiple phase transitions by changing the anisotropy $J'/J$, 
including the topological ones accompanied by the changes of $N_{\rm m}$. 
In this section, we discuss the topological aspects in detail, by tracing the real-space positions of 
the monopoles and antimonopoles corresponding to the hedgehogs and antihedgehogs, respectively, 
as well as the emergent magnetic field from the noncoplanar spin configurations.

Figures~\ref{fig:fig4} and \ref{fig:fig5} display the real-space pictures of the monopoles and antimonopoles  
(magenta and cyan balls) for the $3Q$ and $4Q$ cases, respectively.
The positions are determined by calculating the monopole charge in Eq.~(\ref{eq:monopole_charge}). 
We also plot ${\bm \chi}_{{\bf r}_l}$ in Eq.~(\ref{eq:chi}) by arrows to show the flow of the emergent magnetic field. 
For better visibility, we display the arrows whose norm is larger than a certain threshold in each case.

First, we discuss the $3Q$ case in Fig.~\ref{fig:fig4}. 
In the $2Q$-VC under the hard-axis anisotropy, although there are no monopoles and antimonopoles, 
weak but nonzero ${\bm \chi}_{{\bf r}_l}$ appears locally around the antivortex cores, 
as exemplified in Fig.~\ref{fig:fig4}(a) for $J'/J=0.875$.
${\bm \chi}_{{\bf r}_l}$ are all parallel to the [001] direction, while they form staggered flow around each core. 
While increasing $J'/J$ above $\simeq 0.875$, the system undergoes a phase transition from the $2Q$-VC to the $3Q$-HL. 
At this phase transition, monopoles and antimonopoles are created in pair from the flows of ${\bm \chi}_{{\bf r}_l}$, 
and accordingly, the flows change qualitatively; ${\bm \chi}_{{\bf r}_l}$ starts to flow from the monopoles (sources) 
to the antimonopoles (sinks), as shown in Fig.~\ref{fig:fig4}(b). 
While further increasing $J'/J$, the monopoles and antimonopoles change their relative positions with keeping their $z$ coordinates, 
and for the isotropic case with $J'/J=1.0$ shown in Fig.~\ref{fig:fig4}(c), 
they form spiral structures in all the $x$, $y$, and $z$ directions~\cite{Kanazawa2016,Okumura2020}.  
When we introduce weak easy-axis anisotropy, the monopoles and antimonopoles move again with keeping the $z$ coordinates and approach each other as shown in Fig.~\ref{fig:fig4}(d), and eventually annihilate in pair at the transition 
to the $3Q$ state at $J/J'\simeq 0.9$.  
Even in this topologically trivial state with no monopoles and antimonopoles, there remains nonzero flow of 
${\bm \chi}_{{\bf r}_l}$, especially near the pair annihilation points, as shown in Fig.~\ref{fig:fig4}(e) for $J/J'=0.875$.

Next, we discuss the 4$Q$ case in Fig.~\ref{fig:fig5}.
In the 3$Q$-HL state under the hard-axis anisotropy, there are four pairs of monopoles and antimonopoles 
in the magnetic unit cell, as exemplified in Fig.~\ref{fig:fig5}(a) for $J'/J=0.75$. 
Note that the volume of the magnetic unit cell is four times smaller than the whole cube in Fig.~\ref{fig:fig5}(a), 
as mentioned above. 
${\bm \chi}_{{\bf r}_l}$ distribute to interconnect the monopoles and antimonopoles. 
While increasing $J'/J$ above $\simeq 0.75$, the system undergoes a phase transition from the 3$Q$-HL 
to the 4$Q$-HL. 
At this phase transition, monopoles and antimonopoles connected by the green dotted lines in Fig.~\ref{fig:fig5}(a) 
remain and the others disappear. 
The remaining ones change their relative positions with keeping the sizes and shapes of the two green cubes, as exemplified 
in Fig.~\ref{fig:fig5}(b) for $J'/J=0.775$. 
While further increasing $J'/J$, the monopoles and antimonopoles come to form the body-centered-cubic lattice 
for the isotropic case with $J'/J=1.0$ shown in Fig.~\ref{fig:fig5}(c).
When we introduce weak easy-axis anisotropy, the monopoles of one cube and antimonopoles of the other cube 
approach each other, and ${\bm \chi}_{{\bf r}_l}$ becomes relatively strong between them, as shown in Fig.~\ref{fig:fig5}(d) 
for $J/J'=0.775$.
By further decreasing $J/J'$, the system undergoes the phase transition from the 4$Q$-HL 
to the 4$Q$ state caused by the pair annihilation of the remaining monopoles and antimonopoles.
Similar to the 3$Q$ state in Fig.~\ref{fig:fig4}(e), even after the monopoles and antimonopoles vanish, 
nonzero flow of ${\bm \chi}_{{\bf r}_l}$ remains near the pair annihilation points, 
as exemplified in Fig.~\ref{fig:fig5}(e) for $J/J'=0.75$.

\section{Summary \label{sec:sec5}}

In conclusion, we have theoretically investigated the control of the spin textures and 
their topological properties by changing the spatial anisotropy in the long-range interactions generated by 
the itinerant nature of electrons. 
We showed that the two types of the three-dimensional HLs, which are stable in the spatially isotropic cases, 
change their dimensionality and topology for the hard- and easy-axis anisotropy along one of the helical directions. 
In the case of the HL composed of three orthogonal helices ($3Q$-HL), we found that the HL is 
changed into a two-dimensional vortex crystal ($2Q$-VC) for the hard-axis anisotropy, 
while into a topologically nontrivial $3Q$ state, another $2Q$-VC, and finally, 
a one-dimensional helix (1$Q$-H) as increasing the easy-axis anisotropy. 
On the other hand, in the case of the HL composed of four tetrahedral helices ($4Q$-HL), we found that 
the HL turns into a $3Q$-HL for the hard-axis anisotropy, while into a topologically trivial $4Q$ state and 
a $1Q$-H as increasing the easy-axis anisotropy. 
We showed that the transitions between the topologically nontrivial and trivial states are characterized by pair annihilation 
of the monopoles and antimonopoles. 
Furthermore, we clarified the systematic changes of the topological properties of these spin textures, focusing on 
the real-space positions of the monopoles and antimonopoles, and the flows of the emergent magnetic field defined by 
the scalar spin chirality. 
Our results indicate that the spatial anisotropy in chiral magnets drives not only the magnetic phase transitions 
with dimensional reduction but also the topological properties from the Berry phase mechanism. 

Experimentally, two types of the HLs as well as SkL are realized in MnSi$_{1-x}$Ge$_x$~\cite{Tokunaga2015,Fujishiro2019}. 
Although the microscopic origin is under debate, a scenario was proposed based on the itinerant nature of electrons included 
in the present model in Eq.~(\ref{eq:JKmodel}) with isotropic interactions~\cite{Okumura2020}. 
In this model, the anisotropy would be introduced through the change of the electronic state of itinerant electrons, and hence, our results suggest that 
the interesting transitions found above might be caused, e.g., by external pressure and chemical substitution, which modifies the electronic band structure. 
Such experimental studies as well as further sophisticated theories based on realistic model parameters 
are left for future work.

\begin{acknowledgments}
This research was supported by Grant-in-Aid for Scientific Research Grants Numbers 
JP18K03447, JP19H05822 and JP19H05825, and JST CREST (JP-MJCR18T2), and 
the Chirality Research Center in Hiroshima University and 
JSPS Core-to-Core Program, Advanced Research Networks.
K.S. was supported by the Program for Leading Graduate Schools (MERIT-WINGS). 
S.O. was supported by JSPS through the research fellowship for young scientists. 
Parts of the numerical calculations were performed in the supercomputing systems in ISSP, the University of Tokyo.

\end{acknowledgments}

\bibliography{VC}

\end{document}